\documentclass[12pt]{article}
\usepackage{fullpage}
\usepackage{cite}
\usepackage{amstext,amsfonts,amsbsy,eucal,amssymb}
\usepackage{epsfig}
\usepackage{psfrag}

\newcounter{definition}

\def\thefigure{\arabic{section}.\arabic{figure}}
\def\theequation{\thesection.\arabic{equation}}
\def\appendix{
  \setcounter{section}{0}
  \setcounter{subsection}{0}
  \par
  \def\thesection{Appendix \Alph{section}}
  \def\theequation{\Alph{section}.\arabic{equation}}
  \def\thefigure{\Alph{section}.\arabic{figure}}
}
\makeatletter
\def\fnum@figure{Fig. \thefigure}
\@addtoreset{equation}{section}
\@addtoreset{figure}{section}
\makeatother

\date{
March 12,2001
}
\title{Graphical representation  of  the partition function \\
for a 1-D $\delta$-function Bose gas} 
\author{
  Go {\sc Kato}\thanks{kato@monet.phys.s.u-tokyo.ac.jp}\, and  
  Miki {\sc Wadati}\thanks{wadati@phys.s.u-tokyo.ac.jp} \\
  Department of Physics, Graduate School of Science,\\
  University of Tokyo,\\
  Hongo 7-3-1, Bunkyo-ku, Tokyo 113-0033, Japan.
}

\begin{document}
\maketitle
\setlength{\baselineskip}{1.8em}
\begin{abstract}
    One-dimensional repulsive $\delta$-function bose system is
    studied.
    By only using the Bethe ansatz equation, $n$-particle partition
    functions are exactly calculated. From this expression for the
    $n$-particle partition function, the $n$-particle cluster integral
    is derived.  The results completely agree with those of the
    thermal Bethe ansatz (TBA). This directly proves the validity of
    the TBA. The theory of partitions and graphs is used to simplify
    the discussion.
\end{abstract}

\newpage
\section{Introduction}
We study statistical mechanics of a one-dimensional gas of Bose
particles interacting through a repulsive delta function potential. The
Hamiltonian for the system with $n$ particles reads
\begin{eqnarray}
H_n&=&-\sum_{i=1}^{n}
          \frac{\partial^2}{\partial x_i^2}
         +2c \sum_{i > j}\delta \left(x_i-x_j\right)
       .
\label{eq:Hamiltonian}
\end{eqnarray}
Throughout the paper, we set $\hbar = 1$, $2m = 1$ and assume the
potential is repulsive, $c > 0$.

The eigenvalues and the eigenstates are obtained by the Bethe ansatz (BA)
method \cite{lieb,bethe}, and the quantum inverse scattering
method (QISM)~\cite{creamer_1,creamer_2,thacker_1,thacker_2,Korepin,wadati_c}.
As a periodic boundary condition, the Bethe ansatz (BA) equation is
derived.  Using the BA equation, thermodynamic quantities are calculated
by the thermal Bethe ansatz (TBA)~\cite{wadati_b,yang}. In TBA,
an ``interpretation'' of the particle density and the state density
enables us to define the entropy of $n$-particle system in the
thermodynamic limit.

With the quantum Gelfand-Levitan equation \cite{creamer_1,wadati_c}, the field
operator is expressed as a series of the terms each of which is a
product of creation and annihilation operators in the scattering data space. The grand partition
function is written as a field operator \cite{thacker_1}. Creamer,
Thacker and Wilkinson~\cite{creamer_2} calculated the grand partition
function using creation and annihilation operators of this system, but
the analysis involves a delicate regularization that
 $2\pi\delta(0)$ is replaced by the volume
$L$.

To be rigorous, it is desirable to examine these results by a
different approach. In this paper, we present a method to calculate
the thermodynamical quantities only by use of the BA equation which is
derived exactly as a periodic boundary condition from both QISM and BA
method.
We calculate the $n$-particle partition function
and evaluate the $n$-particle cluster integral.

The paper is organized as follows. The
$n$-particle partition function and the $n$-particle cluster integral
are derived with a method, which we call a direct method, in section
2.
This method is shown explicitly for $n=3$.
In section 3, we consider the $n$-particle case, and reformulate
the results in graphical expressions. The last section is devoted to
concluding remarks and discussions.  To avoid 
complexities, the details of a mathematical proof are summarized in Appendices A.

\section{Direct method}
We assume a finite system size $L$ and the periodic boundary
condition.  It is known that the total energy $E$ and the wave numbers
$k_j$ of the system (\ref{eq:Hamiltonian}) are determined by the
following relations,
\begin{eqnarray}
E    &  =& \sum_{j=1}^{n}k_j^2,
\label{eq:bethe_1}\\
k_jL&=&2\pi m_j+\sum_{j'\neq j}\Delta\left(k_{j'}-k_{j }\right),
\label{eq:bethe_2}
\label{eq:boundary_condition}
\end{eqnarray}
where $m_j$ are integers or half-integers, and $\Delta(k)$ is the phase
shift of two-particle scattering,
\begin{eqnarray}
m_j&\in&\left\{
\begin{array}{ll}
\mathbb{Z}& \makebox{if $N={\rm odd}$}\\
\mathbb{Z}+\frac12& \makebox{if $N={\rm even}$},\\
\end{array} \right.
\\&&
m_j<m_{j+1},
\label{eq:condition_of_n}\\
\Delta\left(k\right)&=&2\arctan\left(\frac k c\right),
\label{eq:bethe_3}
\label{eq:define_delta}
\\
&&-\pi<\Delta\left(k\right)<\pi. \nonumber
\end{eqnarray}
Eq.(\ref{eq:boundary_condition}) is called the Bethe ansatz (BA) equation
.
Main objects to be calculated are the $n$-particle partition function
$Z_n$;
\begin{eqnarray}
Z_n&=&{\rm Tr}e^{-\beta H_n},\quad\quad\beta\quad=\quad1/k_BT,
\end{eqnarray}
and the $n$-particle cluster integral $b_n$;
\begin{eqnarray}
\sum_{n\geq1}b_nz^n&=&\log\left( \sum_{n\geq 0}Z_nz^n\right),
\label{eq:define_bn}
\end{eqnarray}
where $z=e^{\beta\mu}$ with the chemical potential $\mu$. By
definition, $Z_0=1$ and we simply have
\begin{eqnarray}
b_1=L^{-1}Z_1=\int\frac{dk}{2\pi}e^{-\beta k^2}.
\label{eq:define_Z1}
\end{eqnarray}

We explain a direct method to evaluate the partition function for 3-particle
case.  To be explicit, the total energy is
\begin{eqnarray}
   E
 &=&
   k_1^2+k_2^2+k_3^2,
\label{eq:energy_3}
\end{eqnarray}
and the BA equation is
\begin{eqnarray}
   k_1L
 &=&
   2\pi m_1
  +\Delta\left(k_{2}-k_{1}\right)
  +\Delta\left(k_{3}-k_{1}\right)\nonumber\\
   k_2L
 &=&
   2\pi m_2
  +\Delta\left(k_{1}-k_{2}\right)
  +\Delta\left(k_{3}-k_{2}\right)\nonumber\\
   k_3L
 &=&
   2\pi m_3
  +\Delta\left(k_{1}-k_{3}\right)
  +\Delta\left(k_{2}-k_{3}\right),\\
&&m_1<m_2<m_3\in\mathbb{Z}.
\label{eq:m_condition_3}
\end{eqnarray}
By use of  these
relations~(\ref{eq:energy_3})$\sim$(\ref{eq:m_condition_3}), we can calculate the partition function $Z_3$ as follows,
\begin{eqnarray}
  Z_3
& = & 
  \sum_{m_1<m_2<m_3} e^{-\beta\left(k_1^2+k_2^2+k_3^2\right)}
\nonumber\\& = & 
  \frac16\sum_{m_1,m_2,m_3} e^{-\beta\left(k_1^2+k_2^2+k_3^2\right)}
 -\frac12\sum_{m_1,m_2=m_3} e^{-\beta\left(k_1^2+k_2^2+k_3^2\right)}
 +\frac13\sum_{m_1=m_2=m_3} e^{-\beta\left(k_1^2+k_2^2+k_3^2\right)}
\nonumber\\& = & 
  \frac16\sum_{\tilde m_1,\tilde m_2,\tilde m_3}\int dm_1dm_2dm_3
  e^{
    -\beta\left(k_1^2+k_2^2+k_3^2\right)
    +2\pi i\left(m_1\tilde m_1+m_2\tilde m_2+m_3\tilde m_3\right)
  }
\nonumber\\&&{}
 -\frac12\sum_{\tilde m'_1,\tilde m'_2}\int dm'_1,dm'_2
 e^{
   -\beta\left(2k_1^{\prime2}+k_2^{\prime2}\right)
   +2\pi i\left(m'_1\tilde m'_1+m'_2\tilde m'_2\right)
 }
\nonumber\\&&{}
 +\frac13\sum_{\tilde m''} \int dm''
 e^{-\beta\left(3k^2\right)+2\pi i m''\tilde m''}
\nonumber\\& = & 
  \frac16\sum_{\tilde m_1,\tilde m_2,\tilde m_3}\int dk_1dk_2dk_3
   \left|\frac{\partial m}{\partial k}\right|
  e^{
    -\beta\left(k_1^2+k_2^2+k_3^2\right)
    +iL\left(k_1\tilde m_1+k_2\tilde m_2+k_3\tilde m_3\right)
  }\nonumber\\&&{}\times
  \left(\frac{k_1-k_2+ic}{k_1-k_2-ic}\right)^{\tilde m_2-\tilde m_1}
  \left(\frac{k_2-k_3+ic}{k_2-k_3-ic}\right)^{\tilde m_3-\tilde m_2}
  \left(\frac{k_3-k_1+ic}{k_3-k_1-ic}\right)^{\tilde m_1-\tilde m_3}
\nonumber\\&&{}
 -\frac12\sum_{\tilde m'_1,\tilde m'_2}\int dk'_1,dk'_2
   \left|\frac{\partial m'}{\partial k'}\right|
 e^{
   -\beta\left(2k_1^{\prime2}+k_2^{\prime2}\right)
   + iL\left(k'_1\tilde m'_1+k'_2\tilde m'_2\right)
 }
 \left(\frac{k'_1-k'_2+ic}{k'_1-k'_2-ic}\right)^{\tilde m_2-2\tilde m_1}
\nonumber\\&&{}
 +\frac13\sum_{\tilde m''} \int dk''
   \frac{\partial m''}{\partial k''}
 e^{-\beta\left(3k^2\right)+iLk''\tilde m''},
\label{eq:Zn_finite}
\end{eqnarray}
where $|\partial m/\partial k|$, $|\partial m'/\partial k'|$ and
$\partial m''/\partial k''$ are the Jacobians to be explained shortly.  We
have written explicitly all the steps of calculations which are common
to those for general $n$ \cite{go_1,Go_b,Go_c}. In the second
equality, we use a symmetry of the BA equation with respect to the
exchange $m_i,k_i\leftrightarrow m_{i'}, k_{i'}$, in the third
equality, we apply the Poisson's summation formula, and in the last
equality, we change variables of integration from $m$ to $k$.

The relation between $k$ and $m$ is defined by the BA equation.
In (\ref{eq:Zn_finite}), $k'$, $m'$, $k''$ and $m''$ are related as follows,
\begin{eqnarray}
   k'_1L
 &=&
   2\pi m'_1
  +2\Delta\left(k'_2-k'_1\right)\nonumber\\
   k'_2L
 &=&
   2\pi m'_2
  + \Delta\left(k'_1-k'_2\right)
\\
   k''L
 &=&
   2\pi m''.
\end{eqnarray}
Thus, the Jacobians, $|\partial m/\partial k|$, $|\partial
m'/\partial k'|$ and $\partial m''/\partial k''$, are given by
\begin{eqnarray}
   \left(2\pi\right)^3
   \left|\frac{\partial m}{\partial k}\right|
 &=&
    L^3
   +2L^2K\left(k_1-k_2\right)
   +2L^2K\left(k_2-k_3\right)
   +2L^2K\left(k_3-k_1\right)\nonumber\\{}&&
   +3LK\left(k_2-k_1\right)K\left(k_3-k_1\right)
   +3LK\left(k_1-k_2\right)K\left(k_3-k_2\right)\nonumber\\{}&&
   +3LK\left(k_1-k_3\right)K\left(k_2-k_3\right),
\\
   \left(2\pi\right)^2
   \left|\frac{\partial m'}{\partial k'}\right|
 &=&
    L^2
   +3LK\left(k_1-k_2\right),
\\
   \frac{\partial m''}{\partial k''}
 &=&
   \frac 1{2\pi}L,
\end{eqnarray}
where
\begin{eqnarray}
  K\left(k\right)
&\equiv&
\frac{d\Delta\left(k\right)}{dk}
\quad=\quad\frac{2c}{k^2+c^2}.
\end{eqnarray}

It is readily shown that all terms except $\tilde m_i=0$, $\tilde
m'_i=0$ and $\tilde m''=0$ in (\ref{eq:Zn_finite}) exponentially decay
as $L$ gets large.
Although the discussion may include these decaying terms, we hereafter
consider the expressions
in the thermodynamic
limit.  Then, $Z_3$ becomes
\begin{eqnarray}
Z_3&=& 
  \frac16\int \frac{dk_1}{2\pi}\frac{dk_2}{2\pi}\frac{dk_3}{2\pi}
    \left(L^3
   +6L^2K\left(k_1-k_2\right)
   +9LK\left(k_2-k_1\right)K\left(k_3-k_1\right)\right)
  e^{
    -\beta\left(k_1^2+k_2^2+k_3^2\right)
  }\nonumber\\&&{}
 -\frac12\int \frac{dk_1'}{2\pi}\frac{dk_2'}{2\pi}
    \left(L^2 + 3LK\left(k_1-k_2\right)\right)
 e^{
   -\beta\left(2k_1^{\prime2}+k_3^{\prime2}\right)
 }
\nonumber\\&&{}
 +\frac13\int \frac{dk''}{2\pi}L
 e^{-\beta\left(3k^2\right)}.
\label{eq:Z3_infty}
\end{eqnarray}
It is much easier to show
\begin{eqnarray}
Z_2&=& 
  \frac12\int \frac{dk_1'}{2\pi}\frac{dk_2'}{2\pi}
    \left(L^2 + 2LK\left(k_1-k_2\right)\right)
 e^{
   -\beta\left(k_1^{\prime2}+k_3^{\prime2}\right)
 }
\nonumber\\&&{}
 -\frac12\int \frac{dk''}{2\pi}L
 e^{-\beta\left(2k^{\prime\prime2}\right)}.
\label{eq:Z2_infty}
\end{eqnarray}
From (\ref{eq:define_Z1}), (\ref{eq:Z3_infty}) and (\ref{eq:Z2_infty}),
the cluster integral $b_3$ is given by
\begin{eqnarray}
b_3&=&\frac1L\left(Z_3-Z_2Z_1+\frac13Z_1^3\right)\nonumber\\
&= & 
  \frac32\int \frac{dk_1}{2\pi}\frac{dk_2}{2\pi}\frac{dk_3}{2\pi}
  K\left(k_2-k_1\right)K\left(k_3-k_1\right)
  e^{-\beta\left(k_1^2+k_2^2+k_3^2\right)}\nonumber\\&&{}
 -\frac32\int \frac{dk_1'}{2\pi}\frac{dk_2'}{2\pi}
  K\left(k_1-k_2\right)
 e^{-\beta\left(2k_1^{\prime2}+k_3^{\prime2}\right)}
\nonumber\\&&{}
 +\frac13\int \frac{dk''}{2\pi}
 e^{-\beta\left(3k^2\right)}.
\label{eq:b_3}
\end{eqnarray}
Thanks to the effective interaction $K(k)$, $b_3$ consists of only
$3$ terms. In the next section, we show that $Z_n$ and $b_n$ can be
calculated in the same way.

\section{Partition function and graphs}
\subsection{Partition function}
\label{subsec:partition_function}
In order to present a general structure of the partition function, we
need to explain some terminology in the theory of partition \cite{andrews}; the partially ordered set (poset, for short).

\begin{figure}
\begin{center}
\begin{psfrags}
      \psfrag{a}{$123=\hat1_3$}
      \psfrag{b}{$1$-$23$}
      \psfrag{c}{$3$-$12$}
      \psfrag{d}{$2$-$13$}
      \psfrag{e}{$1$-$2$-$3=\hat0_3$}
  \psfig{file=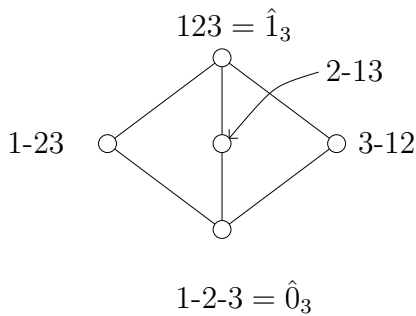 , scale = 0.15}
\end{psfrags}
\end{center}
\caption{
  Vertices of this graph are the elements of $\Pi_3$, and if $x<y$ then
  $y$ is drawn ``above'' $x$ (i.e., with a higher vertical
  coordinate). It is called the {\it Hasse diagram}~\cite{andrews} of $\Pi_3$.}
\label{fig:Pi_3}
\end{figure}

Let $\Pi(S)$ denote a set of all partitions of a finite set $S$.
In what follows, $[n]$ means a set $\{1,2,\cdots,n\}$, and write
$\Pi_n$ for $\Pi([n])$. We define partially order in $\Pi(S)$ by {\it
  refinement}, that is, define $x\leq y\in\Pi(S)$ if every block of
$x$ is contained in a block of $y$. For example, if $x\in\Pi_9$ has
blocks 137-2-46-58-9 and $y\in\Pi_9$ has blocks 13467-2589,
then $x\leq y$. Special elements in $\Pi\left(S\right)$ are $\hat0_S$ and
$\hat1_S$ such that $x\geq\hat0_S$ and $x\leq\hat1_S$ for all
$x\in\Pi(S)$. We write $\hat0_n$ and $\hat1_n$ for $\hat0_{[n]}$ and
$\hat1_{[n]}$.  Fig.\ref{fig:Pi_3} shows the partially ordered
elements in $\Pi_3$, that is , $1$-$2$-$3<1$-$23$, $2$-$13$, $3$-$12<123$.
In addition, we define M\"obius function of $\Pi(S)$ inductively as follows,
\begin{eqnarray}
\mu\left(x,x\right)&=&1
\makebox[3cm]{}
\makebox{for all $x\in \Pi(S)$},
\nonumber\\
\mu\left(x,y\right)&=&-\sum_{x\leq z<y}\mu\left(x,z\right)
\quad\makebox{for all $x<y$ in $\Pi(S)$}.
\label{eq:define_mu}
\end{eqnarray}

By use of those definitions, we can show some useful relations. For
brevity, $\mathbb{N}$ and $\mathbb{C}$ stand respectively nonnegative integers and complex numbers. $(1)$. Let $\hat f,\hat g: \Pi(S)\rightarrow \mathbb{C}$,
then,
\begin{eqnarray}
\hat g\left(x\right)=\sum_{y\geq x}\hat f\left(y\right),
\quad\quad\makebox{for all }x\in\Pi\left(S\right),
\nonumber
\end{eqnarray}
if and only if
\begin{eqnarray}
\hat f\left(x\right)=\sum_{y\geq x}\mu\left(x,y\right)\hat g\left(y\right),
\quad\quad\makebox{for all }x\in\Pi\left(S\right).
\label{eq:Moebius_inversion_formula}
\end{eqnarray}
This is called  M\"obius inversion formula.
$(2)$. Let $\hat g,\hat J: \Pi(S)\rightarrow \mathbb{C}$ and
 $f, h: \mathbb{N} \rightarrow \mathbb{C}$. If
\begin{eqnarray}
    \sum_{n\geq0}h\left(n\right)\frac{u^n}{n!}
  &=&
    \exp\left( \sum_{n\geq1}f\left(n\right)\frac{u^n}{n!}\right),
\label{eq:poset_condition_1}
\\
    h\left(N_S\right)
  &=&
    \sum_{x\in\Pi\left(S\right)}\mu\left(\hat0_S,x\right)\hat g\left(x\right),
\label{eq:poset_condition_2}
\\
    \hat g\left(x\right)
  &=&
    \sum_{\xi\in\Pi\left(x\right)}\prod_{y\in\xi}\hat J\left(y\right),
\label{eq:poset_condition_3}
\end{eqnarray}
then
\begin{eqnarray}
    f\left(n\right)
  &=&
    \sum_{x\in\Pi_n}\mu\left(\hat0_n,x\right)\hat J\left(x\right).
\label{eq:poset_result_1}
\end{eqnarray}
We prove this relation in \ref{sec:poset_proof}.

With these two relations, we can derive the partition function $Z_n$
and the cluster integral $b_n$ in a compact way.  First we define $f, h: \mathbb{N}
\rightarrow \mathbb{C}$ and $\hat h,\hat g: \Pi(S)\rightarrow
\mathbb{C}$,
\begin{eqnarray}
    f\left(n\right)
  &\equiv&
    n!Lb_n,
\\
    h\left(n\right)
  &\equiv&
    n!Z_n,
\label{eq:definition_h_n}\\
    \hat h\left(x\right)
  &\equiv&
    \sum_{m'_        1 \neq              \cdots\neq m'_        l }
    e^{-\beta\left(k_   1 ^2+          \cdots+k_   n ^2\right)},
\label{eq:definition_hat_h}\\
    \hat g\left(x\right)
  &\equiv&
    \sum_{m'_        1 ,               \cdots, m'_        l }
    e^{-\beta\left(k_   1 ^2+          \cdots+k_   n ^2\right)},
\label{eq:definition_hat_g}
\end{eqnarray}
where
\begin{eqnarray}
  x  \in  \Pi\left(S\right)
,\quad\quad
  \sigma_i\ni s_j   
  \;\Rightarrow\;
  m'_        i    =   m_   j
,\quad\quad
  x   =   \left\{\sigma_1,         \cdots,\sigma_l\right\}
,\quad\quad
  S   =   \left\{ s    _1,         \cdots, s    _n\right\}.
\end{eqnarray}
Recall that $m$ and $k$ are related by the  BA equation.
We see that Eq.(\ref{eq:poset_condition_1}) holds from the definition of cluster integral $b_n$, Eq.(\ref{eq:define_bn}).
From the definition (\ref{eq:definition_h_n})$\sim$(\ref{eq:definition_hat_g}), it is easy to show that
\begin{eqnarray}
\hat g\left(x\right)&=&\sum_{y\geq x}\hat h\left(y\right),
\label{eq:tmp_g_from_h} \\
\hat h\left(\hat0_n\right)&=&h\left(n\right).
\end{eqnarray}
Due to the M\"obius inversion formula (\ref{eq:Moebius_inversion_formula}), (\ref{eq:tmp_g_from_h}) is equivalent to 
\begin{eqnarray}
\hat h\left(x\right)&=&\sum_{y\geq x}\mu\left(x,y\right)\hat g\left(y\right).
\end{eqnarray}
Substituting $\hat0_n$ for $x$ in this equation, we obtain
\begin{eqnarray}
h\left(n\right)
&=&
\sum_{y\in\Pi_n}\mu\left(\hat0_n,y\right)\hat g\left(y\right). 
\label{eq:poset_result_2}
\end{eqnarray}
This is nothing but the condition (\ref{eq:poset_condition_2}).

\begin{figure}
\begin{center}
\begin{psfrags}
      \psfrag{a}{\large{$\sigma_1$}}
      \psfrag{b}{\large{$\sigma_2$}}
      \psfrag{c}{\large{$\sigma_3$}}
  \psfig{file=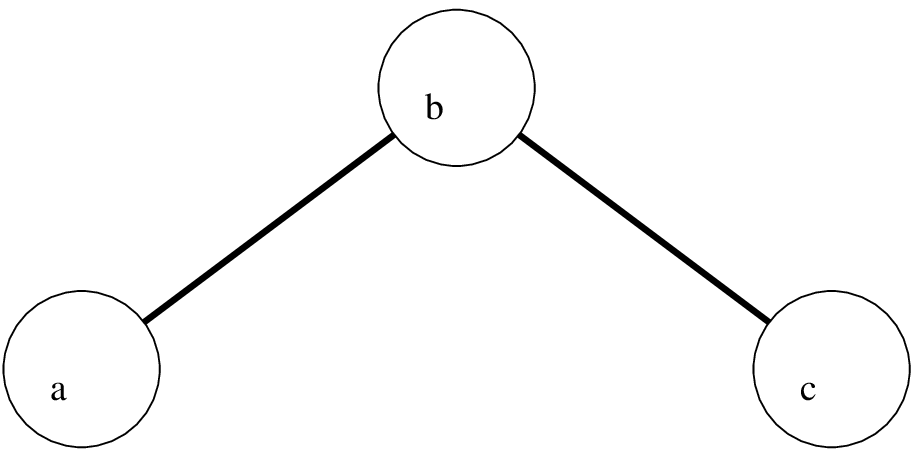 , scale = 0.5}
\end{psfrags}
\makebox[0.5cm]{}
\begin{psfrags}
      \psfrag{a}{\large{$\sigma_2$}}
      \psfrag{b}{\large{$\sigma_3$}}
      \psfrag{c}{\large{$\sigma_1$}}
  \psfig{file=eps/tree111.eps , scale = 0.5}
\end{psfrags}
\makebox[0.5cm]{}
\begin{psfrags}
      \psfrag{a}{\large{$\sigma_3$}}
      \psfrag{b}{\large{$\sigma_1$}}
      \psfrag{c}{\large{$\sigma_2$}}
      \psfrag{d}{\large{vertex}}
      \psfrag{e}{\large{branch}}
  \psfig{file=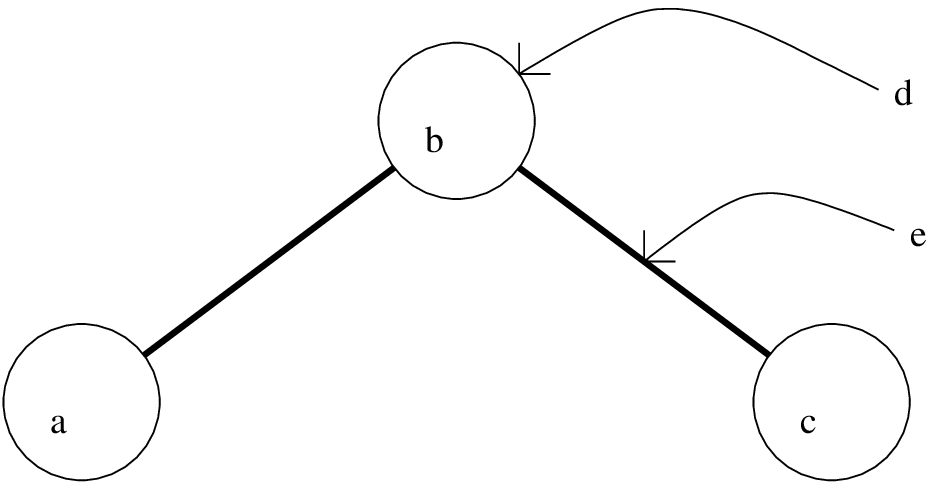 , scale = 0.5}
\end{psfrags}
\end{center}
\caption{A set of undirected trees, $\mathcal{V}(\{\sigma_1,\sigma_2,\sigma_3\})$, contains three elements}
\label{fig:V_x}
\end{figure}

As we have mentioned in the previous section,
 we do not include exponentially decaying  terms. Then, (\ref{eq:definition_hat_g}) is written as
\begin{eqnarray}
    \hat g\left(x\right)
  &=&
    \int \prod_{\sigma\in x}dm'_\sigma
    e^{-\beta\left(k_1^2+k_2^2+\cdots+k_n^2\right)}\\
  &=&
    \int \prod_{\sigma\in x} dk'_\sigma 
    \left|\frac{\partial m'}{\partial k'}\right|
    e^{-\beta\left(\sum_{\sigma\in x} N_\sigma k_\sigma^{\prime 2}\right)}.
\label{eq:define_hat_g}
\end{eqnarray}
The transformation matrix and the Jacobian are given as follows;
\begin{eqnarray}
\frac{\partial m'_\sigma}{\partial k'_{\sigma'}}
&=&
\frac1{2\pi}\times\left\{
\begin{array}{ll}
L+\sum_{\sigma'' \neq \sigma}N_{\sigma''}K\left(k'_{\sigma''}-k'_\sigma\right)
&
\makebox{if $\sigma=\sigma'$}
\label{eq:Jacobi_det_n}\\
-N_{\sigma'}K\left(k'_{\sigma'}-k'_\sigma\right)&
\makebox{if $\sigma\neq {\sigma'}$}
\end{array}
\right.\\
&&\sigma,\sigma',\sigma''\in x\nonumber\\
\left|\frac{\partial m'}{\partial k'}\right|\!
&=&\!\!\!\!
\sum_{\xi\in\Pi\left(x\right)}\!
\prod_{y\in\xi}\frac{L}{\left(2\pi\right)^{N_y}}\!\!
\left(\sum_{\sigma\in y}N_\sigma\right)\!\!\!
\sum_{t\in\mathcal{V}\left(y\right)}\!\!\!
\left(\prod_{\sigma\in y}N_\sigma^{n\left(\sigma,t\right)-1}\right)\!\!\!\!
\left(\prod_{b\in \Xi\left(t\right)}K\left(k'_{\sigma_1\left(b\right)}-k'_{\sigma_2\left(b\right)}\right)\right)\!\!.
\label{eq:Jacobian_n}
\end{eqnarray}
We explain notations in (\ref{eq:Jacobi_det_n}) and
(\ref{eq:Jacobian_n}). The number of elements in a set $\sigma$ is
denoted by $N_\sigma$. $\mathcal{V}\left(S\right)$ denotes a set of
all undirected {\it trees} with a {\it vertex} set $S$. For instance, all
the elements in $\mathcal{V}(\{\sigma_1,\sigma_2,\sigma_3\})$ are
depicted in Fig.\ref{fig:V_x}. We call a connection of two vertices a
{\it branch}. $\Xi\left(t\right)$ denotes a set of all branches
contained in a tree $t$. $n(\sigma,t)$ is the number of branches
with which the vertex $\sigma$ are connected in the tree $t$. 
$\sigma_1(b)$ and $\sigma_2(b)$ denote two end-vertices connected by a branch
$b$.

A remaining task is a consistency with (\ref{eq:poset_condition_3}). We define a function $\hat J: \Pi(S)\rightarrow \mathbb{C}$ by
\begin{eqnarray}
    \hat J\left(x\right)
  &\equiv&
    \int \prod_{\sigma\in x} dk'_\sigma 
    \left|\frac{\partial m'}{\partial k'}\right|_c
    e^{-\beta\left(\sum_{\sigma\in x} N_\sigma k_\sigma^{\prime 2}\right)},
\end{eqnarray}
where
\begin{eqnarray}
\left|\frac{\partial m'}{\partial k'}\right|_c
&\equiv&
\frac{L}{\left(2\pi\right)^{N_x}}
\left(\sum_{\sigma\in x}N_\sigma\right)
\sum_{t\in\mathcal{V}\left(x\right)}
\left(\prod_{\sigma\in x}N_\sigma^{n\left(\sigma,t\right)-1}\right)
\left(\prod_{b\in\Xi\left(t\right)}K\left(k'_{\sigma_1\left(b\right)}-k'_{\sigma_2\left(b\right)}\right)\right).
\label{eq:Jacobian_c}
\end{eqnarray}
We can change the expression of the Jacobian $|\partial m'/\partial
k'|$ from (\ref{eq:Jacobian_n}) into a sum of forests, that is, sets of trees.
 On the other hand, the right hand side
of (\ref{eq:Jacobian_c}) is a sum of connected forests, that is, trees. Therefore, we put a subscript $c$ as
$|{\partial m'}/{\partial k'}|_c$.
With this definition, we see that the condition (\ref{eq:poset_condition_3}) follows from (\ref{eq:define_hat_g}).

In this way, three conditions
(\ref{eq:poset_condition_1})$\sim$(\ref{eq:poset_condition_3}) are
shown to be satisfied, which indicates that Eq.(\ref{eq:poset_result_1}) holds.
We write Eq.(\ref{eq:poset_result_1}) explicitly,
\begin{eqnarray}
  b_n
 &=&
  \frac1{n!L}\sum_{x\in\Pi_n}\mu\left(\hat0_n,x\right)
  \int \prod_{\sigma\in x} dk'_\sigma 
  \left|\frac{\partial m'}{\partial k'}\right|_c
  e^{-\beta\left(\sum_{\sigma\in x} N_\sigma k_\sigma^{\prime 2}\right)}.
\label{eq:bn_using_poset}
\end{eqnarray}
In fact, (\ref{eq:poset_result_2}) is equivalent to 
\begin{eqnarray}
  Z_n
 &=&
  \frac1{n!}\sum_{x\in\Pi_n}\mu\left(\hat0_n,x\right)
  \int \prod_{\sigma\in x} dk'_\sigma 
  \left|\frac{\partial m'}{\partial k'}\right|
  e^{-\beta\left(\sum_{\sigma\in x} N_\sigma k_\sigma^{\prime 2}\right)}.
\label{eq:Zn_using_poset}
\end{eqnarray}
The Jacobians in (\ref{eq:bn_using_poset}) and
(\ref{eq:Zn_using_poset}) are respectively (\ref{eq:Jacobian_c}) and
(\ref{eq:Jacobian_n}).
 Explicit form of
$\mu(\hat0_n,x)$ can be derived from (\ref{eq:define_mu}),
\begin{eqnarray}
  \mu\left(\hat0_n,x\right)
&=&
\prod_{\sigma\in x}\left(-1\right)^{N_\sigma-1}\left(N_\sigma-1\right)!.
\end{eqnarray}
It is easily  shown \cite{Go_b,Go_c} that the cluster integrals $b_n$ agree with those derived from
the thermal Bethe ansatz (TBA).  It is also possible to
derive the integral equation in TBA from (\ref{eq:bn_using_poset}) \cite{Go_c}.

We give two remarks here. First, the Jacobian $|\partial m'/\partial
k'|$ is essentially the inner product of the Bethe wave
functions~\cite{Korepin}. Second, the cluster integrals consist of only a finite number of terms.

\subsection{Graph representation}
\begin{figure}
\begin{center}
\begin{psfrags}
      \psfrag{a}{$l$}
  \psfig{file=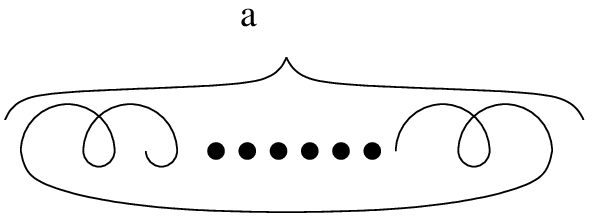 , scale = 0.5}
\end{psfrags}
\end{center}
\caption{$l$-toron.}
\label{fig:l_toron}
\end{figure}

\begin{figure}
\begin{center}
\begin{psfrags}
  \psfig{file=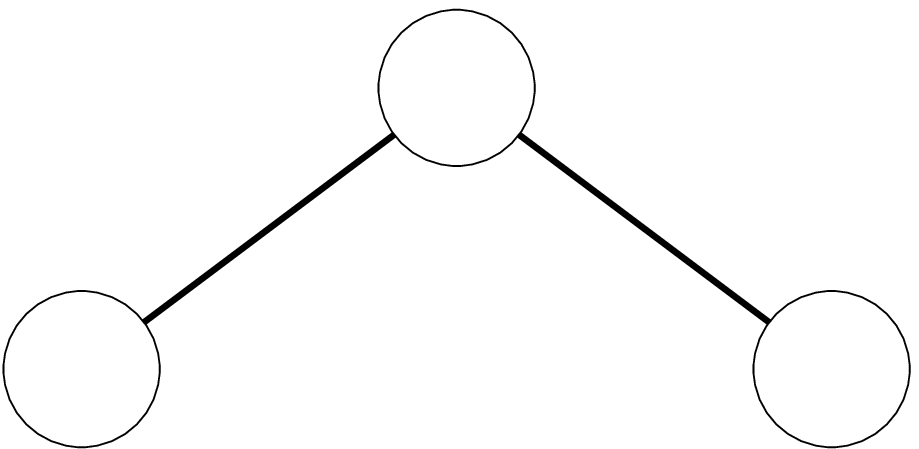 , scale = 0.5}
\end{psfrags}
\makebox[0.5cm]{}
\begin{psfrags}
  \psfig{file=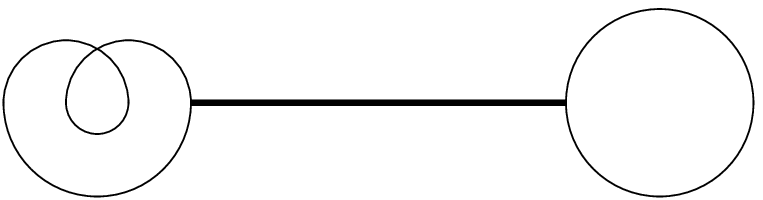 , scale = 0.5}
\end{psfrags}
\makebox[0.5cm]{}
\begin{psfrags}
  \psfig{file=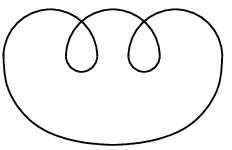 , scale = 0.5}
\end{psfrags}
\end{center}
\caption{A set of the trees, $\mathcal{V}_3$, contains three elements}
\label{fig:V_3}
\end{figure}

\begin{figure}
\begin{center}
\begin{psfrags}
  \psfig{file=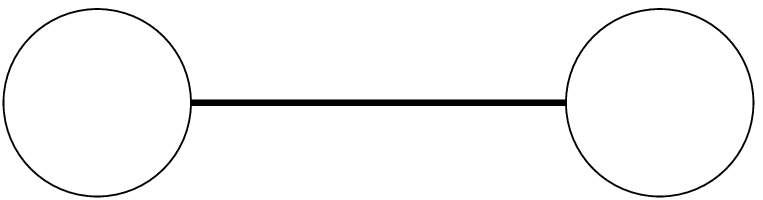 , scale = 0.5}
\end{psfrags}
\begin{psfrags}
  \psfig{file=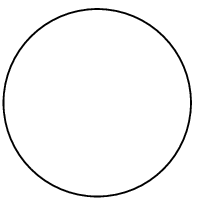 , scale = 0.5}
\end{psfrags}
\makebox[1.3cm]{}
\begin{psfrags}
  \psfig{file=eps/tree1n.eps , scale = 0.5}
\end{psfrags}
\begin{psfrags}
  \psfig{file=eps/tree1n.eps , scale = 0.5}
\end{psfrags}
\begin{psfrags}
  \psfig{file=eps/tree1n.eps , scale = 0.5}
\end{psfrags}
\makebox[1.7cm]{}
\begin{psfrags}
  \psfig{file=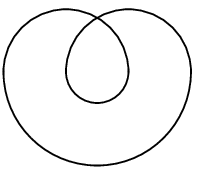 , scale = 0.5}
\end{psfrags}
\begin{psfrags}
  \psfig{file=eps/tree1n.eps , scale = 0.5}
\end{psfrags}
\\
\begin{psfrags}
  \psfig{file=eps/tree111n.eps , scale = 0.5}
\end{psfrags}
\makebox[1cm]{}
\begin{psfrags}
  \psfig{file=eps/tree21n.eps , scale = 0.5}
\end{psfrags}
\makebox[2cm]{}
\begin{psfrags}
  \psfig{file=eps/tree3n.eps , scale = 0.5}
\end{psfrags}
\end{center}
\caption{A set of the forests, $\mathcal{W}_3$. Three graphs in the first line are not trees because the vertices are not connected on partially connected by branch(es).}
\label{fig:W_3}
\end{figure}

We further develop a graphical representation of the cluster integral
and the partition function.  We draw an $l$ times rolled coil for a
Boltzmann weight $e^{-\beta lk^2}$ (Fig.\ref{fig:l_toron}). We call it
$l$-toron following Montroll and Ward~\cite{montroll,Isihara}.
The {\it tree} consists of a toron or torons connected by
branches. The {\it forest} consists of the trees. We denote by
$\mathcal{V}_n$ a set of all the trees
 which satisfy the
following two conditions: $(1)$ all the vertices of the tree are made
of torons, and $(2)$ the sum of rolled number of torons composing the
tree is $n$.  Fig.\ref{fig:V_3} shows all the elements in
$\mathcal{V}_3$.
$\mathcal{W}_n$ is a set of all the  forests which satisfy the
following two conditions: $(1)$ all the vertices of the forest are
made of torons, and $(2)$ the sum of rolled number of torons composing
the forest is $n$. Simply, $\mathcal{V}_n$ is a subset of
$\mathcal{W}_n$. Fig.\ref{fig:W_3} exhibits all the elements in
$\mathcal{W}_3$.

In terms of these terminologies, we may rewrite 
(\ref{eq:Zn_using_poset}) and (\ref{eq:bn_using_poset}) as
\begin{eqnarray}
  Z_n
 &=&
  \sum_{f\in\mathcal{W}_n}\frac{Sym\left(f\right)}{n!}
  \prod_{t \makebox{ in the forest $f$}} \mathcal{S}\left(t\right),
\label{eq:Zn_using_toron}
\\
  b_n
 &=&
  \sum_{t\in\mathcal{V}_n}\frac{Sym\left(t\right)}{n!L}
  \mathcal{S}\left(t\right).
\label{eq:bn_using_toron}
\end{eqnarray} 
Here, $Sym\left(t\right)$ and $Sym\left(f\right)$ indicate symmetrical
factors of graphs.  In the case $t\in\mathcal{V}_n$ or $f\in\mathcal{W}_n$,
$Sym\left(t\right)$ or $Sym\left(f\right)$ means the number of
different ways in which a set $\{1,\cdots,n\}$ can be distributed to
all the vertices of $t$ or $f$ at a time, on the condition that $l$
elements are placed in a vertex made of $l$-toron. For example
\begin{eqnarray}
Sym\left(
\begin{psfrags}
  \psfig{file=eps/tree111n.eps , scale = 0.1}
\end{psfrags}
\makebox[0.1cm]{}
\begin{psfrags}
  \psfig{file=eps/tree1n.eps , scale = 0.1}
\end{psfrags}
\right)
&=&12,\\
Sym\left(
\begin{psfrags}
  \psfig{file=eps/tree21n.eps , scale = 0.1}
\end{psfrags}
\right)
&=&3,
\label{eq:b3_Symterm21}\\
Sym\left(
\begin{psfrags}
  \psfig{file=eps/tree1n.eps , scale = 0.1}
\end{psfrags}
\makebox[0.1cm]{}
\begin{psfrags}
  \psfig{file=eps/tree1n.eps , scale = 0.1}
\end{psfrags}
\makebox[0.1cm]{}
\begin{psfrags}
  \psfig{file=eps/tree1n.eps , scale = 0.1}
\end{psfrags}
\right)
&=&1.
\end{eqnarray} 
From (\ref{eq:bn_using_poset}) and (\ref{eq:Zn_using_poset}), we can
show that $\mathcal{S}(t)$ in (\ref{eq:Zn_using_toron}) and (\ref{eq:bn_using_toron}) is given as follows,
\begin{eqnarray}
  \mathcal{S}\left(t\right)&\equiv&L
  \sum_{\omega}N_\omega
  \left(
    \prod_{\omega}
    \left(-1\right)^{N_\omega-1}\left(N_\omega-1\right)!N_\omega^{n\left(\omega,t\right)-1}
  \right)
\nonumber\\&&{}\times
  \int \prod_{\omega} \frac{dk_\omega}{2\pi}
  \left(\prod_{b\in\Xi\left(t\right)}
  K\left(k_{\sigma_1\left(b\right)}-k_{\sigma_2\left(b\right)}\right)\right)
  e^{-\beta\left(\sum_{\omega} N_\omega k_\omega\right)},
\end{eqnarray}
where $N_\omega$ denotes the rolled number of toron $\omega$, and
$\sum_\omega$ (or $\prod_\omega$) denotes a sum (or a product) with respect
to all the toron $\omega$ in the tree $t$.  For example,
\begin{eqnarray}
  \mathcal{S}\left(
    \begin{psfrags}
      \psfig{file=eps/tree21n.eps , scale = 0.1}
    \end{psfrags}
  \right)
  &=&L\times\left(2+1\right)\times\left(\left(-1\right)^11!2^0\times\left(-1\right)^00!1^0\right)
  \nonumber\\&&{}
  \times\int \frac{dk_{\omega_1}}{2\pi} \frac{dk_{\omega_2}}{2\pi}
  \left(K\left( k_{\omega_1}-k_{\omega_2}\right)\right)
  e^{-\beta\left(2k_{\omega_1}+k_{\omega_2}\right)}.
\label{eq:b3_Sterm21}
\end{eqnarray}
Substitution of (\ref{eq:b3_Symterm21}) and (\ref{eq:b3_Sterm21}) gives the second term in (\ref{eq:b_3}).

As examples, we list a graphical representation of the  cluster integrals $b_1\sim b_4$,
\begin{eqnarray}
b_1&=&
\begin{psfrags}
  \psfig{file=eps/tree1n.eps , scale = 0.2}
\end{psfrags}
\\
b_2&=&
\begin{psfrags}
  \psfig{file=eps/tree11n.eps , scale = 0.2}
\end{psfrags}
+
\begin{psfrags}
  \psfig{file=eps/tree2n.eps , scale = 0.2}
\end{psfrags}
\\
b_3&=&
\begin{psfrags}
  \psfig{file=eps/tree111n.eps , scale = 0.2}
\end{psfrags}
+
\begin{psfrags}
  \psfig{file=eps/tree21n.eps , scale = 0.2}
\end{psfrags}
+
\begin{psfrags}
  \psfig{file=eps/tree3n.eps , scale = 0.2}
\end{psfrags}
\\
b_4&=&
\begin{psfrags}
  \psfig{file=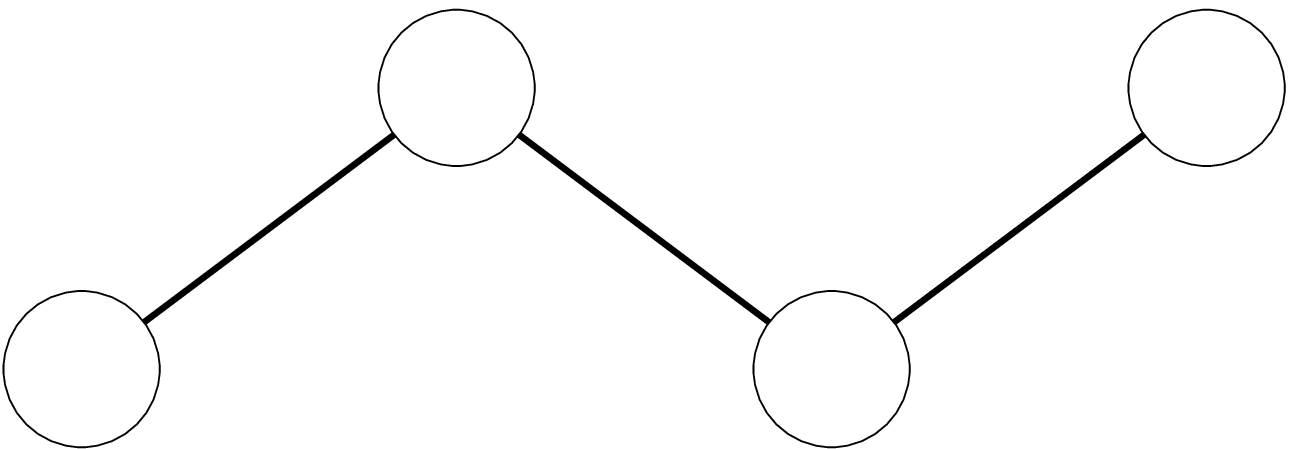 , scale = 0.18}
\end{psfrags}
+
\begin{psfrags}
  \psfig{file=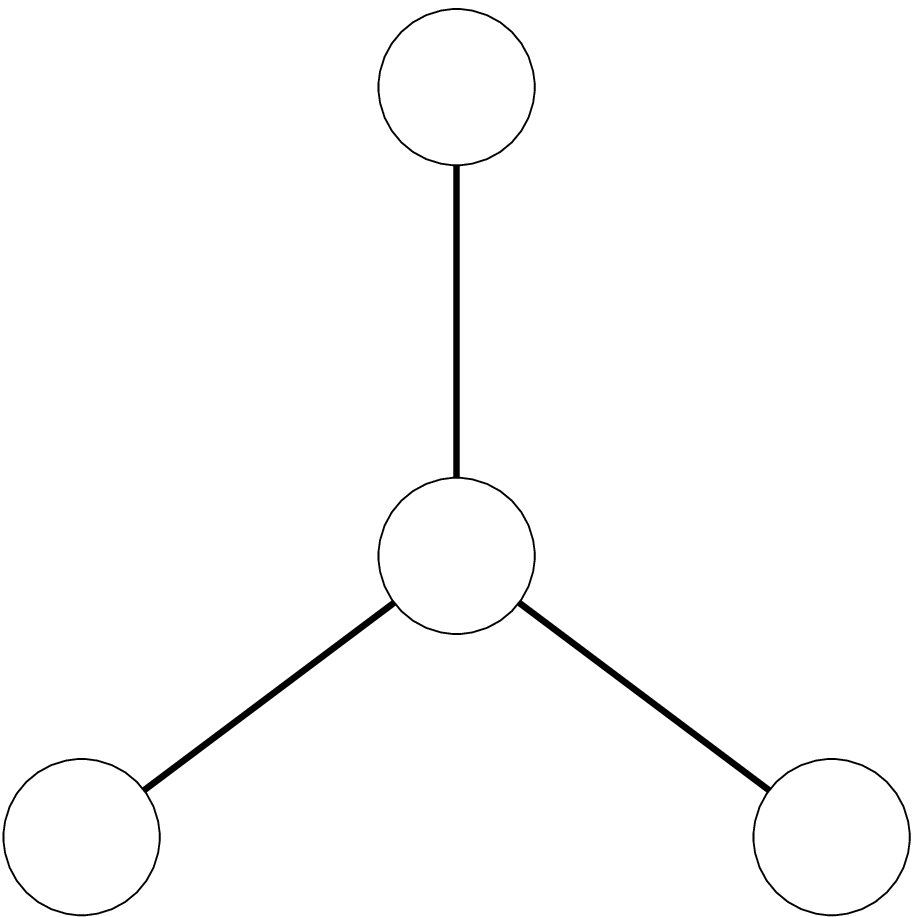 , scale = 0.18}
\end{psfrags}
+
\begin{psfrags}
  \psfig{file=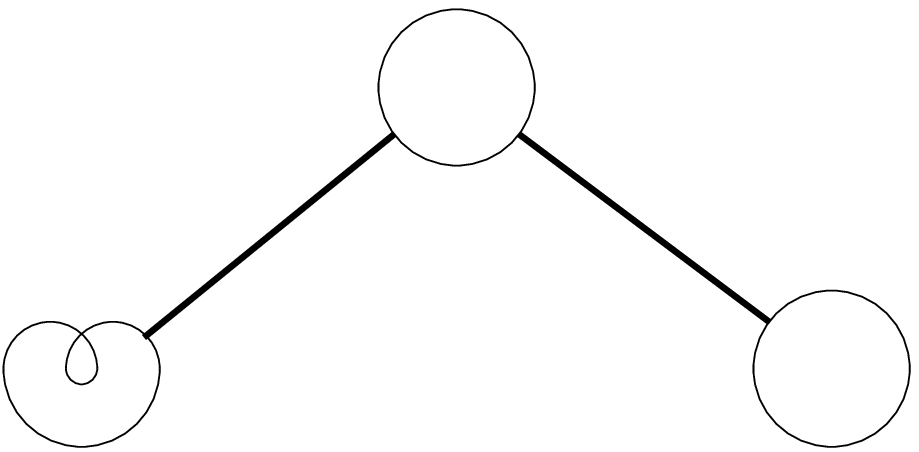 , scale = 0.18}
\end{psfrags}
+
\begin{psfrags}
  \psfig{file=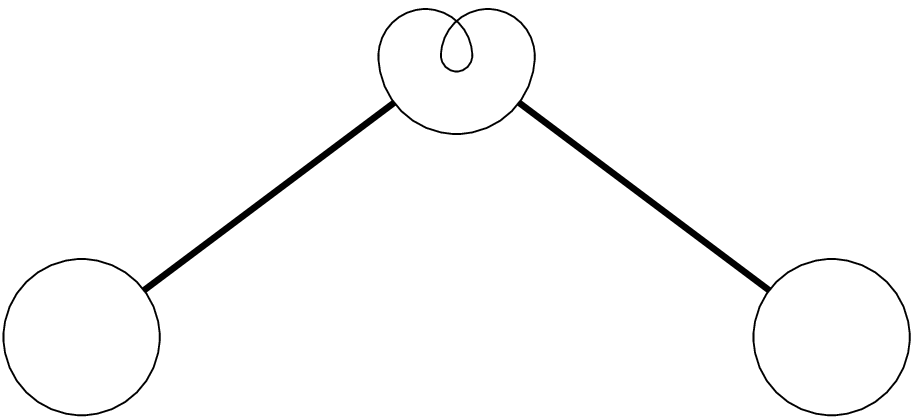 , scale = 0.18}
\end{psfrags}
\nonumber\\&&{}
+
\begin{psfrags}
  \psfig{file=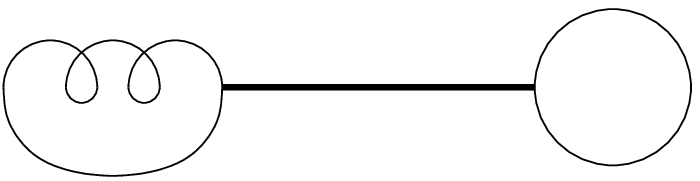 , scale = 0.2}
\end{psfrags}
+
\begin{psfrags}
  \psfig{file=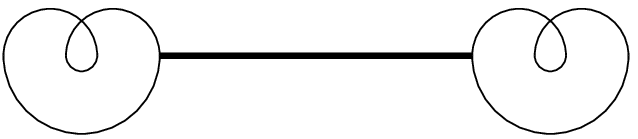 , scale = 0.2}
\end{psfrags}
+
\begin{psfrags}
  \psfig{file=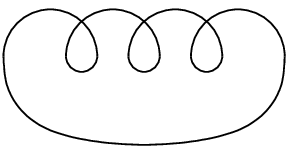 , scale = 0.2}
\end{psfrags}
.\
\end{eqnarray} 

\section{Concluding remarks and discussions}
In this paper, we have studied a one-dimensional $\delta$-function
bose gas, (\ref{eq:Hamiltonian}), and have derived directly the
partition function and the cluster integral, (\ref{eq:Zn_using_poset})
and (\ref{eq:bn_using_poset}), from the Bethe ansatz equation,
(\ref{eq:boundary_condition}). This should be regarded as a proof of the thermal Bethe ansatz
(TBA). The derivation is simplified by use of the partially ordered
set in the theory of partition.

This direct method has some advantages.  First, the method is
rigorous.  In the quantum field theoretic method, calculation is done
at the infinite volume, therefore an interpretation of
$\;2\pi\delta(0)$ as the volume is unavoidable.  In the TBA, it is
necessary to define the $n$-particle entropy. This procedure is based
on an interpretation
of the particle and state densities.
In this sense, both field theoretic method and TBA remained to be
proved. On the other hand,
the direct method is free from such interpretations, and all
calculations are traceable step by step.

Second, the method has a wide applicability.  It should be remarked that
in the direct method all the dependencies on the systems come only from
the BA equation.  In other words, the direct method may have the
generality at least as well as TBA.

We conclude that the direct method will be useful and important in
calculating the thermodynamic quantities in various integrable systems.

\appendix

\section{ A proof of (\ref{eq:poset_result_1})}
\label{sec:poset_proof}
We use the same notation as in subsection
\ref{subsec:partition_function}. It can be shown that,
on condition that $h,f: \mathbb{N}\rightarrow \mathbb{C}$, we have
\begin{eqnarray}
  \sum_{n\geq0}h\left(n\right)\frac{u^n}{n!}
 &=&
  \exp \left(\sum_{n\geq1}f\left(n\right)\frac{u^n}{n!}\right)
\label{eq:assumption_1}
\end{eqnarray}
if and only if
\begin{eqnarray}
  h\left(n\right)
 &=&
  \sum_{x\in\Pi_n}\prod_{\sigma\in x}f\left(N_\sigma\right)
\label{eq:app_a:Eq2}\\
  h\left(0\right)&=&1.
\end{eqnarray}
This is known as the cumulant expansion formula.
Eq.(\ref{eq:app_a:Eq2}) is equivalent to
\begin{eqnarray}
\prod_{\sigma\in x}h\left(N_\sigma\right)
&=&
\sum_{y\leq x}\prod_{\sigma\in y}f\left(N_\sigma\right).
\label{eq:app_a:Eq3}
\end{eqnarray}

For $\hat H,\hat F: \Pi(S)\rightarrow \mathbb{C}$,
the following relation holds,
\begin{eqnarray}
  \hat H\left(x\right)=\sum_{y\leq x}\hat F\left(y\right)
  &\Longleftrightarrow&
  \hat F\left(x\right)=\sum_{y\leq x}\mu\left(x,y\right)\hat H\left(y\right).
\end{eqnarray}
  This is the M\"obius inversion formula, which is a dual form of (\ref{eq:Moebius_inversion_formula}).
  With this formula, Eq.(\ref{eq:app_a:Eq3}) becomes
\begin{eqnarray}
\prod_{\sigma\in x}f\left(N_\sigma\right)
&=&
\sum_{y\leq x}\mu\left(y,x\right)\prod_{\sigma\in y}h\left(N_\sigma\right).
\label{eq:app_a:Eq4}
\end{eqnarray}
Substitution of  $\hat1_n$ for $x$ in  (\ref{eq:app_a:Eq4}) yields
\begin{eqnarray}
f\left(n\right)
&=&
\sum_{x\leq \hat1_n}\mu\left(y,\hat1_n\right)
\prod_{\sigma\in y}h\left(N_\sigma\right).
\end{eqnarray}

We suppose the existence of $\hat g: \Pi(S)\rightarrow \mathbb{C}$ which satisfies
\begin{eqnarray}
h\left(N_S\right)&=&\sum_{x\in\Pi\left(S\right)}\mu\left(\hat0_S,x\right)\hat g\left(x\right).
\label{eq:assumption_2}
\end{eqnarray}
Using this relation, Eq.(\ref{eq:app_a:Eq4}) is written as
\begin{eqnarray}
  f\left(n\right)
  &=&
  \sum_{x\in\Pi_n}\mu\left(x,\hat1_n\right)
  \prod_{\sigma\in x}\sum_{y\in\Pi\left(\sigma\right)}
  \mu\left(\hat0_\sigma,y\right)\hat g\left(y\right)\nonumber\\
  &=&
  \sum_{x\in\Pi_n}\mu\left(\hat0_n,x\right)
  \sum_{\xi\in\Pi\left(x\right)}\mu\left(\xi,\hat1_x\right)
  \prod_{y\in\xi}\hat g\left(y\right).
\label{eq:app_a:Eq5}
\end{eqnarray}
In the second equality, we have used the following two relations,
\begin{eqnarray}
  \sum_{x\in\Pi_n}F\left(N_x\right)
  \prod_{\sigma\in x}\sum_{y\in\Pi\left(\sigma\right)}
  \hat G\left(y\right)
  &=&
  \sum_{x\in\Pi_n}
  \sum_{\xi\in\Pi\left(x\right)}F\left(N_\xi\right)
  \prod_{y\in\xi}\hat G\left(y\right),
\end{eqnarray}
for $F: \mathbb{N}\rightarrow \mathbb{C}$, $\hat G: \Pi(S)\rightarrow
\mathbb{C}$, and
\begin{eqnarray}
  \mu\left(\hat0_n,x\right)
  &=&
  \prod_{\sigma\in x}\mu\left(\hat 0_{N_{\sigma}},\hat 1_{N_{\sigma}}\right),
\end{eqnarray}
for $x \in\Pi_n$.
And, we suppose the existence of a map $\hat J: \Pi(S)\rightarrow \mathbb{C}$ which satisfies
\begin{eqnarray}
\hat g\left(x\right)&=&
\sum_{\xi\in\Pi\left(x\right)}\prod_{y\in\xi}\hat J\left(y\right).
\label{eq:assumption_3}
\end{eqnarray}

Then, a main part of the right hand side of Eq.(\ref{eq:app_a:Eq5}) becomes
\begin{eqnarray}
  \sum_{\xi\in\Pi\left(x\right)}\mu\left(\xi,\hat1_x\right)
  \prod_{y\in\xi}\hat g\left(y\right)
  &=&
  \sum_{\xi\in\Pi\left(x\right)}\mu\left(\xi,\hat1_x\right)
  \prod_{y\in\xi}\sum_{\zeta\in\Pi\left(y\right)}
  \prod_{z\in\zeta}\hat J\left(z\right)
  \nonumber\\&=&
  \sum_{\zeta\in\Pi\left(x\right)}
  \left(\sum_{\lambda\in\Pi\left(\zeta\right)}
    \mu\left(\lambda,\hat1_\zeta\right)
  \right)
  \prod_{z\in\zeta}\hat J\left(z\right)
  \nonumber\\&=&
  \hat J\left(x\right).
\label{eq:app_a:Eq6}
\end{eqnarray}
The third equality is due to  the following relation,
\begin{eqnarray}
  \sum_{x\in\Pi\left(S\right)}\mu\left(x,\hat1_S\right)&=&\delta\left(1,N_S\right).
\end{eqnarray}
By use of the relation (\ref{eq:app_a:Eq6}), (\ref{eq:app_a:Eq5}) becomes
\begin{eqnarray}
f\left(n\right)=\sum_{x\in\Pi_n}\mu\left(\hat0_n,x\right)\hat J\left(x\right),
\label{eq:app_a:Eq7}
\end{eqnarray}
which is (\ref{eq:poset_result_1}). To summarize, using
(\ref{eq:assumption_1}), (\ref{eq:assumption_2}) and
(\ref{eq:assumption_3}), we have proved (\ref{eq:poset_result_1}).

\end{document}